 \documentclass[aps,prl,notitlepage,superscriptaddress,twocolumn ]{revtex4}
\usepackage{graphicx,subfigure,epsfig}
\usepackage{hyperref}
\usepackage{dcolumn}
\usepackage{amssymb}
\usepackage{times}
\usepackage{amsmath}
\usepackage{amsfonts}
\usepackage{mathrsfs}
\usepackage{setspace}
\usepackage{latexsym}
\usepackage{bbm}
\usepackage{float}
\usepackage{flafter}
\usepackage{bm}
\usepackage{epstopdf}
\usepackage{color}
\usepackage{multirow}
\usepackage{physics} 
\usepackage[export]{adjustbox}

\def \be {\begin{eqnarray}}
\def \ee {\end{eqnarray}}

\begin{document}

\title{Local Probes for Quantum Hall Ferroelectrics and Nematics}

\author{Pok Man Tam\textsuperscript{\textsection}}
\affiliation{Department of Physics and Astronomy, University of Pennsylvania, Philadelphia, PA 19104, USA}

\author{Tongtong Liu\textsuperscript{\textsection}}
\affiliation{Department of Physics, Massachusetts Institute of Technology, Cambridge, MA 02139, USA}

\author{Inti Sodemann}
\affiliation{Max-Planck Institute for the Physics of Complex Systems, D-01187 Dresden, Germany}
\author{Liang Fu}
\affiliation{Department of Physics, Massachusetts Institute of Technology, Cambridge, MA 02139, USA}

\begin{abstract}
Two-dimensional multi-valley electronic systems in which the dispersion of individual pockets has low symmetry give rise to quantum Hall ferroelectric and nematic states in the presence of strong quantising magnetic fields. We investigate local signatures of these states arising near impurities that can be probed via Scanning Tunnelling Microscopy (STM) spectroscopy. For quantum Hall ferroelectrics, we demonstrate a direct relation between the dipole moment measured at impurity bound states and the ideal bulk dipole moment obtained from the modern theory of polarisation. We also study the many-body problem with a single impurity via exact diagonalization and find that near strong impurities non-trivial excitonic state can form with specific features that can be easily identified via STM spectroscopy.

\end{abstract}
\maketitle

\begingroup\renewcommand\thefootnote{\textsection}
\footnotetext{These two authors contributed equally.}
\endgroup

\noindent {\color{blue}\emph{Introduction.}} Recently we have witnessed an explosion of high-quality two-dimensional electronic systems with strongly anisotropic dispersions that can be driven into the quantum Hall regime in the presence of strong magnetic fields \cite{1,22}, such as (111) surface of Bismuth \cite{2,20,21, Agarwal2019}, AlAs heterostructures \cite{3,3b}, PbTe(111) quantum wells \cite{4} and (001) surface of the topological crystalline insulator (TCI) like Sn$_{1-x}$Pb$_x$(Te,Se) \cite{5}. In these systems, at integer fillings of the Landau levels, Coulomb interaction tends to spontaneously break symmetry by forming valley-polarized states \cite{1,6,7,8}, which can be generally divided into nematic or ferrolectric states according to whether or not the Fermi surface of individual valley preserves inversion (or two-fold rotation) symmetry \cite{1}. Advances in STM have made it possible to directly image the shape of Landau orbitals near impurities~\cite{2,20,21,Papic2018}, providing an exciting window into these correlated states.  Evidence of the quantum Hall ferroelectrics has been reported in Bismuth (111) \cite{20}. The surface of SnPb(Te,Se) based TCI's is another promising platform to realise these states \cite{10,11,15,12,13}. 

\begin{figure}[t!]
\begin{center}
\includegraphics[width=0.4\textwidth]{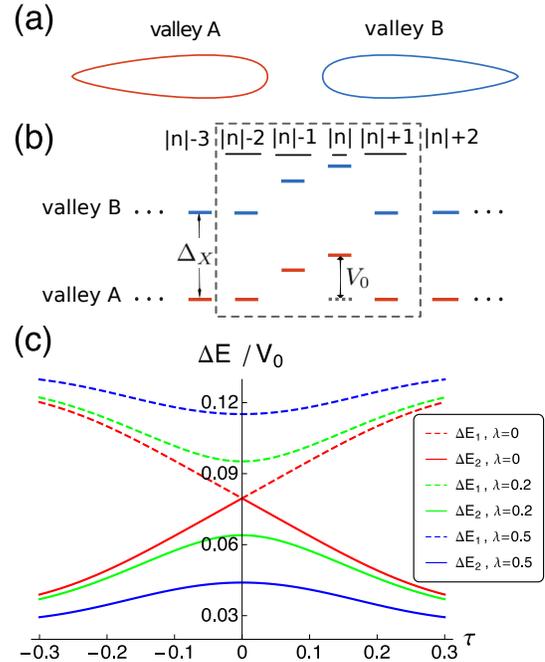}
\end{center}
\par
\renewcommand{\figurename}{FIG.}
\caption{(a) Simplistic illustration of a quantum Hall ferroelectric system. The Fermi surface consists of two valleys related by a two-fold rotation, while individual valley breaks (preserves) two-fold rotation symmetry for the ferroelectric (nematic) state. (b) Schematic of single orbit spectra, for the $n^\text{th}$ Dirac Landau level: upon hybridization only 2 states are perturbed in energy by a delta-function impurity $V_0$~\cite{supp}. The exchange splitting $\Delta_X$  favors valley polarization. (c) Energies, $\Delta E_1$ and $\Delta E_2$, of the two impurity states for the $n=\pm 1$ Dirac Landau level, as a function of the tilt ($\tau$) and mass ($\lambda$) of the Dirac cone.}
    \label{FIG1}
\end{figure}

In this Rapid Communication we investigate the behaviour of quantum Hall ferroelectrics and nematics near short range impurities. One of our goals is to elaborate on how to measure an ``order parameter" for the quantum Hall ferroelectricity. In trivial insulators in which the bulk and the boundary are simultaneously gapped a natural order parameter is the ferroelectric dipole moment, which can be computed from the Berry phase based approach in the modern theory of polarization~\cite{16,17}. In quantum Hall ferroelectrics, although such polarization is well defined in an ideal setting subjected to periodic boundary conditions, it is unclear how to directly measure it due to screening at metallic boundaries. This issue can be resolved by studying states bound to impurities. Indeed, the ideal dipole moment defined by the modern theory of polarization can be related to that of impurity bound states, as we will demonstrate for the case of tilted Dirac cones relevant for the surface of SnPb(Te,Se) based TCI's.

We also study numerically the many-body problem of states near short range impurities by exact diagonalization. As previously discussed~\cite{2,20} the impurities can shift the energy of the occupied state that has a finite amplitude at the impurity location. We have found a new many-body regime where the impurity potential exceeds the exchange energy that attempts to keep the Landau level (LL) completely filled. For repulsive short range impurities, once the impurity potential overcomes this threshold, a state with a quasi-hole bound to the impurity becomes the ground state of the system, and one of the lowest-lying excited states corresponds to a non-trivial inter-valley excitonic state, in which an electron is added to another valley. We will discuss how these new many-body states have clear signatures in STM measurements.

\noindent {\color{blue}\emph{Impurity states for Dirac cones.}} Here we consider a model that is relevant to the (001) surface of SnPb(Te,Se) based TCI's. In these materials, at temperatures below a ferroelectric transition their surface states comprise four Dirac cones, two of which are massive and two massless. Each of the massive/massless pair is degenerate in the presence of time-reversal symmetry~\cite{13}, but under a background magnetic field the degeneracy of the massive pair is no longer protected. The degeneracy of the massless pair will however remain protected by the product of time reversal and a mirror symmetry (see Supplement \cite{supp}). Here we focus on the latter two degnerate valleys. The dispersions generally have a tilt in momentum~\cite{14,15}, which is essential to the ferroelectricity that we describe below. We thus consider the following effective Hamiltonian for the Dirac cone at $\pm \bar{\Lambda}$ (near $\bar{X}$):
\begin{equation}\label{eq0}
H=v_x \sigma_x p_x - v_y \sigma_y p_y \pm \delta v_x p_x+ \Delta\sigma_z ,
\end{equation}
where $\sigma_i$ are Pauli matrices and $\delta v_x$ represents the tilt of the Dirac cone. For generality we have added a mass term, $\Delta \sigma_z$, to the originally massless Dirac cones which is allowed in magnetic fields due to the Zeeman effect, however, in TCI's this coupling has been seen to be negligibly small \cite{5}. In the presence of external magnetic fields Landau levels will form, and we consider a partial filling $\nu=1$ for the resulting 2-fold degenerate valley doublet. The quantum Hall ferroelectric (nematic) state forms when the electrons spontaneously polarize into a single one of these valleys due to interactions ~\cite{1}. Figure~\ref{FIG1}(a) and (b) provide simplistic illustrations of this model.  
\begin{figure}[t!]
    \includegraphics[width=0.51 \textwidth]{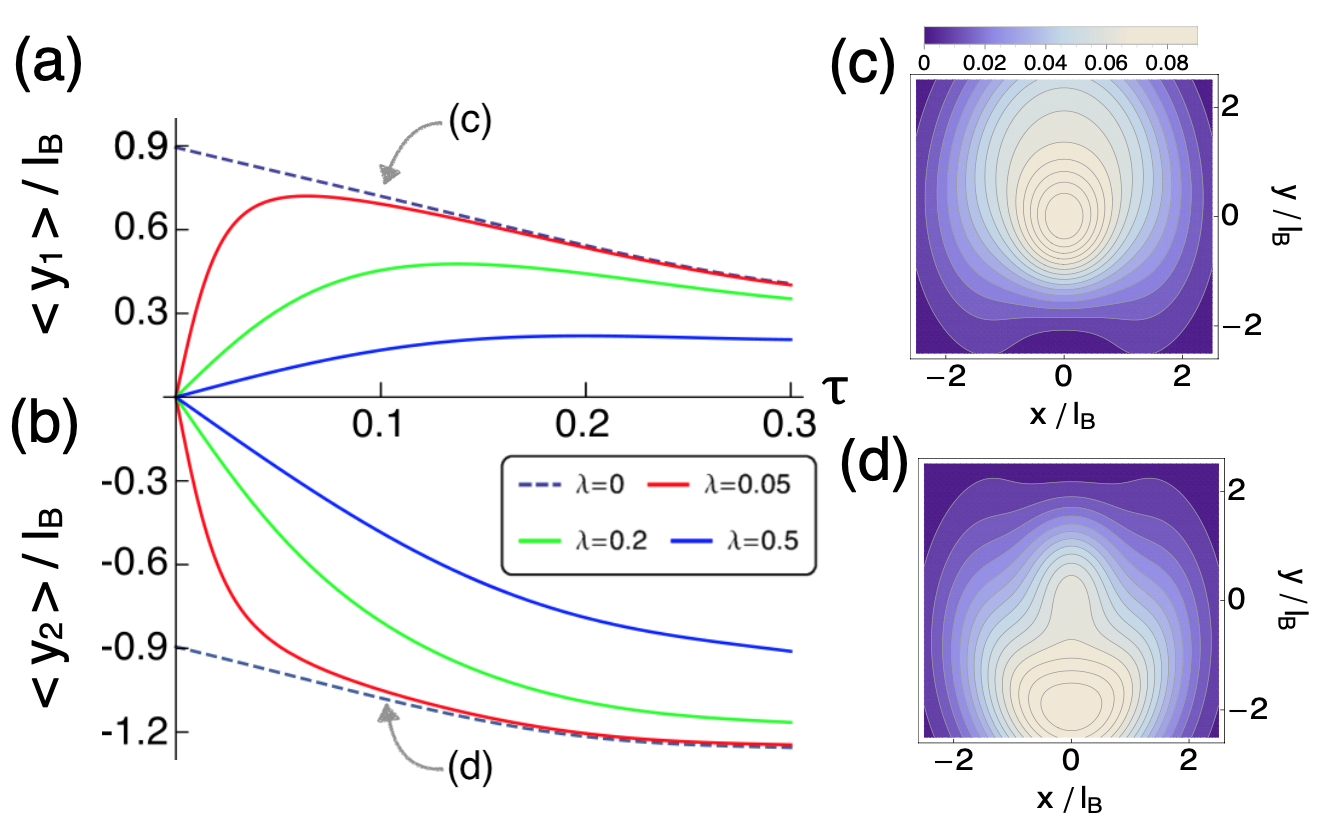}
    \caption{(a), (b): Average position, measured from the impurity site, of the impurity states from the $n=-1$ Dirac LL, as a function of the tilt ($\tau$) and mass ($\lambda$) of the Dirac cone. (c), (d): Spatial probability distribution of the impurity states (for TCI parameters $\tau=0.1$, $\lambda=0$ and $v_x/v_y=1.6$), which can be probed by the tunneling differential conductance in STM.}
    \label{FIG2}
\end{figure}
Inspired by recent STM experiments~\cite{2,20}, we study states near short-range impurities modelled as delta-function potentials~\cite{23}:  
\begin{equation}\label{eq17}
H_{imp} = V_0 l^2_B\; \delta({\bf r}),
\end{equation}
where $l_B=\sqrt{\hbar c/eB}$ is the magnetic length. Assuming that the impurity potential ($V_0$) is smaller than the Landau level spacing, we project the Hamiltonian to the Landau level of interest. Only states with a finite probability at the origin will be affected by the impurity potential. For a parabolic dispersion, there would be a single state per Landau level with non-zero probability at the origin, as demonstrated in the Bismuth experiments~\cite{2}. However, the situation is richer for Dirac Landau levels due to the two-component nature. Some distinctions between the conventional and Dirac Landau levels have been revealed in STM experiments on the surface of topological insulators \cite{twocomponent}, and here we discuss another distinction regarding the impurity state. The wavefunction of the $n^{\text{th}}$ Dirac Landau level in the massless and un-tilted limit (for the general case see the Supplement \cite{supp}) is: 
\begin{equation}\label{eq8}
\psi_{n,m} =
\frac{1}{\sqrt{Z_{n}}}\begin{pmatrix}
\phi_{|n|,m} \\
s_n \phi_{|n|-1,m}
\end{pmatrix},
\end{equation}
\noindent where $n\in \mathbb{Z}$, $s_n={\rm sgn}(n)$ (with $s_0=0$), $Z_n=2^{|s_n|}$, and $\phi_{|n|,m}$ is the wavefunction for a parabolic Landau level in the symmetric gauge with angular momentum $m-|n|$. For the $n=0$ LL only the $m=0$ state would have probability at the origin, however, for $n\neq 0$, two states  with $m=|n|$ and $m=|n|-1$ would have probability at the origin and opposite pseudospins~\cite{18,19}. These two states are exactly degenerate for a massless and un-tilted dispersion, but either of these perturbations produces an energy splitting as illustrated in Fig.~\ref{FIG1}. Thus the impurity states are generically resolvable in STM measurements. In the Supplement~\cite{supp} we demonstrate that these perturbations do not produce extra impurity states, and therefore, only these two states are split from the bulk Landau level and bound to the impurity.  Let us introduce dimensionless parameters to characterise the tilt $\tau \equiv \delta v_x/(2 v_x)$ and the mass $\lambda \equiv \Delta l_B/(\sqrt{2v_xv_y})$. In Sn$_{1-x}$Pb$_x$(Te,Se) these are approximately $\tau=0.1$, $\lambda=0$ (neglecting Zeeman effect) and $v_x/v_y = 1.6$~\cite{supp}. It is therefore justified to use  perturbation theory in $\tau$. The splitting of the two impurity states from the bulk $n=\pm 1$ Landau level, to leading order in $\tau$, are then estimated to be: $\Delta E_1 \approx 0.10 V_0$, $\Delta E_2  \approx 0.06 V_0$. Figure~\ref{FIG2} displays the spatial profile of these two states.\\

\noindent {\color{blue}\emph{Ferroelectric dipole moments.}} In the modern theory of electric polarization \cite{16,17}, the dipole moment of an insulator is computed by adopting periodic boundary conditions. The dipole is computed from the change of the electronic position while varying the Hamiltonian along an adiabatic path in which the bulk gap remains open and that starts from an inversion symmetric reference state. Following this principle a dipole moment for the ferroelectric quantum Hall state was introduced in Ref.~\cite{1}. For tilted Dirac cones, this dipole moment per particle to leading order in the tilt is:
\begin{equation}\label{eq12}
\begin{split}
&{\bf D}_{n} =  {\tilde s}_{n} \sqrt{2}\;\tau\; e \;l_B\;\left(\frac{2\lambda^2+3|n|}{\sqrt{\lambda^2+|n|}}\right)\sqrt{\frac{v_y}{v_x}} \hat{{\bf y}},
\end{split}
\end{equation}

\noindent here $\tilde{s}_n={\rm sgn}(n)$ (with ${\tilde s}_0=1$). Notice that the dipole along the tilt ($x$-axis) vanishes~\cite{1}. The limitation of this definition is that one assumes the charge that flows through the bulk will appear intact at surface, providing a net electric polarization.  However, in an insulating topological phase with a metallic boundary, the latter assumption is unjustified since the surface charge can flow and lead to vanishing macroscopic polarization. Hence it is important to devise alternative diagnostics of inversion asymmetry in topological phases such as the quantum Hall ferroelectrics.

Impurity states, which can be locally probed by STM, offer a resolution. For any given impurity state one can define a dipole moment as the expectation value of the position measured relative to the center of the impurity potential. If the impurity potential is inversion symmetric, this dipole moment serves to characterise the inversion asymmetry of the host state. Figure~\ref{FIG2}(a) and (b) display the average position of the impurity states in tilted Dirac cones as a function of their mass and tilt. Interestingly, the average position is non-analytic, as evidenced by the fact that the limits of $\tau\rightarrow 0$, $\lambda\rightarrow 0$ do not commute in Fig.~\ref{FIG2}. This is a consequence of the fact that in this limit both impurity states are degenerate and hence the expectation values on individual states become ambiguous. However, the sum of the average positions in both impurity states is free from ambiguities and vanishes as $\tau\rightarrow 0$, $\lambda\rightarrow 0$. We therefore introduce the notion of the {\it impurity dipole moment}, ${\bf D}^{\rm imp}$, as the sum of the average position of impurity states $\psi_i$~\footnote{This average coincides with the minus of dipole moment weighted by the charge distribution of the hole that is left in Landau level which can also be directly accessed by STM measurements.}:

\begin{equation}\label{eq1}
{\bf D}^{\rm imp} = e\sum_i\langle \psi_i| {\bf  r}|\psi_i\rangle.
\end{equation}

\noindent To leading order in the tilt ($\tau$) and mass ($\lambda$) of the Dirac cone, we obtained the following relation between the adiabatic bulk dipole moment, in Eq.~\eqref{eq12}, and the impurity dipole moment:

\begin{equation}\label{eq1}
{\bf D}_{n}^{\rm imp} = \frac{2|n|}{3|n|+2\lambda^2} {\bf D}_{n},
\end{equation}

\noindent for the $n^{\text{th}}$ Dirac Landau level in a Dirac cone of mass $\lambda$ (derivation is presented in the Supplement \cite{supp}). This formula summarizes one of the key messages of our study: local measurement of the impurity dipole moment, ${\bf D}^{\rm imp}$, combined with the knowledge of the electronic structure, can be used to probe the bulk adiabatic dipole moment following from the modern theory of polarisation, ${\bf D}$, in a quantum Hall ferroelectric state. 


In the massless limit, i.e. $\lambda\ll \sqrt{|n|}$, the two dipole moments have a simple proportionality relation, ${\bf D}_{n}^{\rm imp} = (2/3) {\bf D}_{n}$. However, a notable difference between these two notions appears in the large mass limit, i.e. $\lambda\gg \sqrt{|n|}$, for which the adiabatic dipole grows linearly with the mass, $|{\bf D}_{n}| \propto \lambda$, whereas $|{\bf D}_{n}^{\rm imp}| \propto 1/\lambda$. This markedly different behavior is a consequence of the approach to the parabolic mass limit as we explain in the Supplement~\cite{supp}.\\

\noindent {\color{blue}\emph{Many-body physics near impurities.}} So far we have largely ignored the role of electron-electron interactions by imagining a large self-consistent exchange field has set in to select a single valley. Next, we will study the many-body problem in the presence of the impurity potential from Eq.~\eqref{eq17} by means of exact diagonalization on a torus. 
We concentrate here on the ferroelectric states where two valleys are described by the tilted massless Dirac cone with the same axis orientation and velocity ratio but opposite tilt. We expect the states at Landau level $n=+3$ to essentially carry over to the case of Bismuth Surfaces~\cite{2,20,21, Agarwal2019}. In the Supplement~\cite{supp}, we also present a nematic model of two valleys with anisotropic masses whose principal axes are rotated by $\pi/2$, as in AlAs quantum wells~\cite{3,3b}, which gives a simpler picture of what we find.

\begin{figure}[t!]
\centering
\includegraphics[width=0.45\textwidth, left]{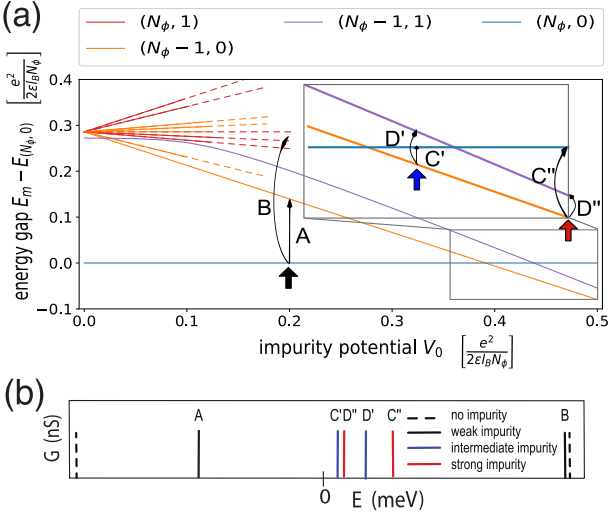}
\caption{(a) Spectra with increasing impurity potential: $(N_\text{A},N_\text{B})$ labels the state with $N_\text{A}$ electrons in valley A and $N_\text{B}$ electrons in valley B. 
Energy is measured relative to that of $(N_{\phi},0)$. Notice that the ground state is changed from $(N_\phi,0)$ to $(N_\phi-1, 0)$ as the repulsive impurity becomes stronger. Here we use $N_\phi=40,\;\tau=0.1,\;v_x/v_y=5$ and $\lambda=0$. (b) Illustration of tunneling peaks measured via STM. The peaks are labeled in correspondence with tunneling processes indicated in the upper panel. For simplicity in (b) we only show one of the two impurity levels that split from the bulk Landau level. The other is visible in panel (a) as a solid-dashed orange line.
}
\label{spectrum}
\end{figure}

In the absence of impurity ($V_0=0$) at the $n=+3$ Dirac LL and partial filling $\nu=1$, the ground state of the system sponaneously polarizes into a single valley and an exchange splitting, $\Delta_X$, between the two valleys develops~\cite{1,7,8}. This is schematically depicted in Fig.~\ref{FIG1}(a) and (b). In the forthcoming discussion we choose the chemical potential to lie exactly in the middle of the charge gap, namely, we add a single particle term to the Hamiltonian so that far away from the impurity the energy to add one electron equals the energy to add one hole. In STM spectra this is satisfied when the two peaks corresponding to the occupied and empty valleys in the Landau level are located symmetrically away from zero bias with no impurity, as illustrated in Fig.~\ref{spectrum}. We assume a sufficiently strong tilt so that the lowest energy charged excitations are not skyrmions~\cite{1}.

We denote the valley polarization of states by a vector $(N_\text{A},N_\text{B})$, where $N_i$ is the number of electrons in valley $i$ ($i=\text{A, B}$). The ground state at $\nu=1$ in the absence of the impurity therefore has polarization $(N_\phi,0)$. The number of orbits in a single valley is taken to be $N_\phi=40$. STM is customarily viewed as a probe of the density of states of the single particle charged excitations, because it requires the removal or injection of electrons from the sample. As we will see, however, near strong impurities, it is possible to use STM to probe excitonic states. For a weak impurity, $V_0\ll \Delta_X$, as the STM tip is brought near the impurity one expects simply a shift of the spectrum by an energy $\sim V_0$, reflecting the local change of energy to remove/add particles as illustrated by peaks $A, B$ in Fig.~\ref{spectrum}. In this regime one encounters excitonic states inside the gap. However, they are {\it invisible} in the STM spectrum because they are {\it neutral} and hence orthogonal to states with added/removed electrons relative to the ground state. 

\begin{figure}[t!]
\centering
\includegraphics[width=0.48\textwidth, center]{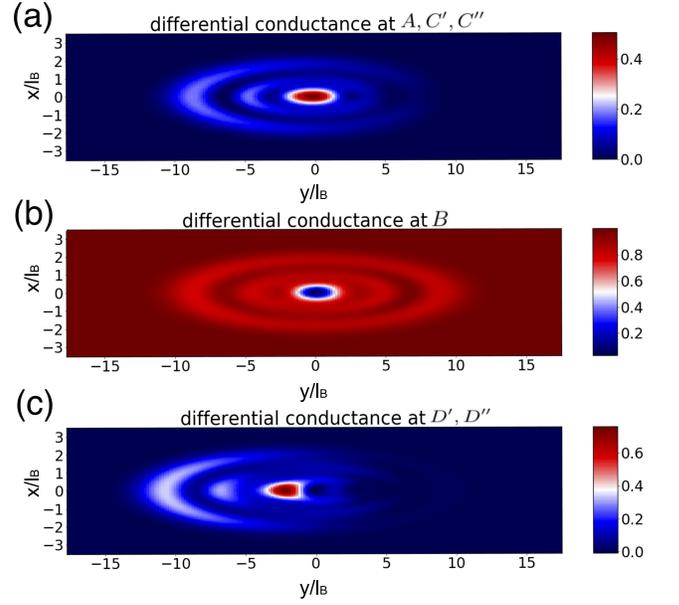}
\caption{The local density of states at energy levels $A,B,C',D',C'',D''$,  which is proportional to the differential conductance obtained by STM measurements. The unit of length is set to be $l_B$. The tilt $\tau=0.1$ and velocity ratio $v_x/v_y=5$ are used.}
\label{tunneling}
\end{figure}

Interestingly, when the impurity potential exceeds a threshold on the order of exchange splitting, the ground state of the system is no longer the fully valley-polarized state, $(N_\phi,0)$, but rather a quasihole state with polarization $(N_\phi-1,0)$~\footnote{Here we describe the behavior for repulsive impurities $V_0>0$, but equivalent statements hold for attractive impurities after performing a particle-hole conjugation $(N_\text{A},N_\text{B})\rightarrow (N_\phi-N_\text{B},N_\phi-N_\text{A})$. Particularly, the quasihole state in the repulsive case is replaced by a quasi-particle state in the attractive case.}, as described in Fig.~\ref{spectrum}(a). This is essentially a local doping of the ground state by removing one electron. Importantly, there appear then two energetically close excited states with quantum numbers $(N_\phi,0)$ and $(N_\phi-1,1)$. These two lowest excited states differ from the ground state by {\it adding} a single electron, and hence will appear as two peaks ($C$ and $D$) at positive bias in the STM spectrum, as shown in Fig.~\ref{spectrum}(b). These two peaks shift sides as $V_0$ increases, when the energy of $(N_\phi,0)$ exceeds $(N_\phi-1,1)$. Experimentally these peaks can be distinguished by probing the respective spatial differential conductance, as detailed in Fig.~\ref{tunneling}. 

The $(N_\phi-1,1)$ state can be viewed as an excitonic state bound to the impurity. Since it differs from the local ground state by one electron its wavefunction can be imaged by STM. The differential conductance of adding an electron in STM is given by the local density of state (LDOS) at energy $\varepsilon$:

\begin{equation}
G({\bf r})\propto\sum_m|\langle \phi_m|\sum_j\left( c^\dagger_{\text{A},j}\phi_{\text{A},j}^*({\bf r})+c^\dagger_{\text{B},j}\phi_{\text{B},j}^*({\bf r})\right)|\phi_0\rangle|^2,
\label{conductance}
\end{equation}

\noindent where $|\phi_0\rangle$ is the lowest energy state. For a weak impurity below the threshold, $|\phi_0\rangle=|N_\phi,0\rangle$. Above the threshold, $|\phi_0\rangle=|N_\phi-1,0\rangle$, which is the hole state created by the impurity. $c^\dagger_{i,j}$ and $\phi_{i,j}$ are the creation operator and single electron wavefunction for an orbit $j$ on valley $i$. $\langle \phi_m|$ is the state with energy $\varepsilon$, the sum over $m$ is taken for all degeneracy. The case of removing an electron follows from Eq.~\eqref{conductance} by replacing $c^\dagger_{i,j}$ and $\phi^*_{i,j}$ with $c_{i,j}$ and $\phi_{i,j}$ respectively.  

Figure~\ref{tunneling} depicts the expected shape of the differential conductance in STM at the energy and impurity indicated in Fig.~\ref{spectrum}. The $B$ peak in the spectroscopy includes multiple nearly degenerate states, here in Fig.~\ref{tunneling} we treat them as degenerate at energy $\varepsilon$ and average over them. The first two panels of Fig.~\ref{tunneling} depict tunneling between a single-hole or electron state and the fully polarized state, which only involves single-body physics; while the last panel is the tunneling between the hole state and the excitonic state, though only reflects the LDOS of valley B with one electron, its shape is modified via the interaction with the hole in valley A. The significant difference between Fig.~\ref{tunneling}(a) and (c) allows for distinguishing this non-trivial excitonic state in STM.\\

\noindent {\color{blue}\emph{Summary.}} We have studied how to locally probe quantum Hall ferroelectric and nematic states near short range impurities. Particularly the impurity dipole moment, which is measurable via STM, is introduced to characterise the degree of inversion asymmetry in quantum Hall ferroelectrics. We have also investigated the many-body problem near strong impurities and found non-trivial excitonic states. These states, though typically invisible in STM near weak impurities, become accessible near strong impurities which change the ground state by locally removing/adding an electron.\\

\begin{acknowledgments}
\noindent {\color{blue}\emph{Acknowledgments.}} We are grateful to Benjamin Feldman, Mallika Randeria and Ali Yazdani for illuminating discussions. T.L. would like to thank Zheng Zhu for helping with the numerical study. This work is supported by DOE Oﬃce of Basic Energy Sciences under Award DE-SC0018945. P.M.T. was in part supported by the Croucher Scholarship for Doctoral Study from the Croucher Foundation.
\end{acknowledgments}

\pagebreak
\widetext

\begin{center}
\textbf{\large Supplementary Materials for ``Local Probes for Quantum Hall Ferroelectrics and Nematics"}\\\text{}
\\
\text{Pok Man Tam, Tongtong Liu, Inti Sodemann, and Liang Fu}
\end{center}

\onecolumngrid
\setcounter{secnumdepth}{2}
\renewcommand{\theequation}{\thesubsection.\arabic{equation}}
\renewcommand{\theHequation}{\theHsubsection.\arabic{equation}}

In this supplementary, we provide more details about the setup of our theoretical and numerical studies. In Sec. \ref{AA}-\ref{D}, we explain how we study integer quantum Hall states on the surface of topological crystalline insulator (TCI), which has Dirac dispersion that is both tilted and massive. In Sec. \ref{AA} we discuss the symmetry on the (001) surface of TCI Sn$_{1-x}$Pb$_x$(Te,Se), and introduce the low-energy $k \cdot p$ model for our subsequent study. The Dirac Landau levels in this system are solved in Sec. \ref{A}, and it is argued in Sec. \ref{B} that in the presence of delta-potential impurity, there are exactly two states per Landau level that are perturbed away in energy. In Sec. \ref{C} we explain the values of parameters adopted in our model. In Sec. \ref{CD} we derive the relation between the local impurity dipole moment and the bulk adiabatic dipole moment, which is quoted in the main text, while in Sec. \ref{D} we distinguish these two notions of dipole and identify the one that can reveal ferroelectricity in our system. In Sec. \ref{E}, we present the setup for carrying out exact diagonalization which leads to the prediction of non-trivial excitonic states near strong impurities. While the experimental signatures of these many-body states in systems with Dirac dispersion have been discussed in the main text, a simpler situation with anisotropic parabolic dispersion (such as in AlAs quantum well) is analyzed in Sec. \ref{F}. 

\subsection{$M\mathcal{T}$ symmetry and its spontaneous breaking}\label{AA}
The low-energy effective Hamiltonian is introduced in Eq. (1) of the main text. Here we describe a symmetry that relates the two valleys under consideration. 

\begin{figure}[H]
\centering
\includegraphics[width=0.25\textwidth, height=4.3cm]{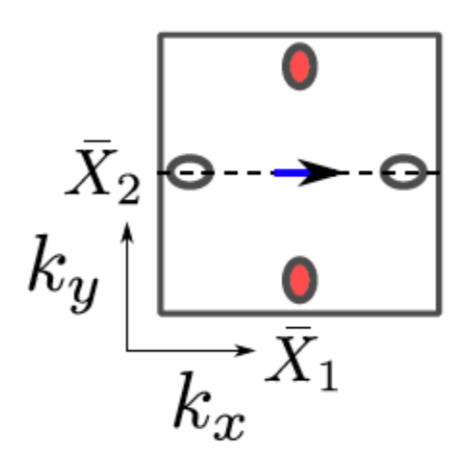}
\caption{Surface Brillouin zone of the (001) surface of TCI. There is a pair of Dirac cones near $\bar{X}_1$ and a pair near $\bar{X}_2$. The blue arrow indicates the ferroelectric distortion. }
\label{BZ}
\end{figure}

We are interested in the (001) surface of TCI. It is known from earlier studies that there is a structural distortion occurring spontaneously at low temperature and breaks all rotation symmetry \cite{supp01}. This distortion can happen along either $[110]$ or $[1\bar{1}0]$ direction, and leads to a ferroelectric polarization. Note that this is \textit{not} the quantum Hall ferroelectricity that we study in this paper. Without loss of generality, let us assume the distortion is along $[110]$ and set up the coordinates such that it is the $x$-direction. Because of the ferroelectric distortion, the (001) surface has only the mirror symmetry $M_y$, which reverses the $y$-direction, and time-reversal $\mathcal{T}$. An illustration of the surface Brillouin zone is shown in Fig. \ref{BZ}. The low-energy effective model consists of four Dirac cones, with a massive degenerate pair near $\bar{X}_1$ and a massless degenerate pair near $\bar{X}_2$ \cite{supp13}. The $2+2$-fold degeneracy is protected by $\mathcal{T}$. 

In the presence of magnetic field, the mirror symmetry $M_y$ and the time-reversal $\mathcal{T}$ are \textit{individually} broken. Still, the product $M_y \mathcal{T}$ remains a symmetry. Notice that $M_y \mathcal{T}$ relates the individual valleys near $\bar{X}_1$ to themselves, while it exchanges opposite valleys near $\bar{X}_2$. Hence, under a magnetic field, such $2+2$-fold degeneracy of Landau levels is explicitly broken down to just the $2$-fold degeneracy for the massless Dirac cones near $\bar{X}_2$. Moreover, such originally massless Dirac cones can now acquire a Zeeman gap, as we explain below.

Let us label the location of the Dirac cones as $\pm \Lambda$. At $+\Lambda$, a two-band model has been constructed before and admits the following form \cite{supp13, supp12}:
\begin{equation}
H_{+\Lambda} = v_x \sigma_y p_x +v_y \sigma_x p_y +\delta v_x p_x +\Delta \sigma_z, 
\end{equation}
Here $\delta v_x$ characterizes the tilt of Dirac cone, which is symmetry-allowed and has been observed experimentally \cite{supp15}. The mirror symmetry $M_y$ acts on the pseudo-spin and momentum as follows:
\begin{equation}
M_y : p_x \rightarrow p_x,\; p_y \rightarrow -p_y,\; \sigma_x \rightarrow -\sigma_x,\; \sigma_y \rightarrow \sigma_y, \; \sigma_z \rightarrow-\sigma_z, 
\end{equation}
while time-reversal $\mathcal{T}$ acts as follows:
\begin{equation}
\mathcal{T}: p_x \rightarrow -p_x,\; p_y \rightarrow -p_y,\; \sigma_x \rightarrow -\sigma_x,\; \sigma_y \rightarrow -\sigma_y, \; \sigma_z \rightarrow-\sigma_z. 
\end{equation}
From these we see that the mass term $\Delta=0$ in the absence of background field, as $M_y$ is a symmetry that relates the cone at $+\Lambda$ to itself but sends $\Delta\sigma_z$ to $-\Delta\sigma_z$. However, in the presence of a magnetic field, $M_y$ is no longer a symmetry, and hence the massless Dirac cone at $+\Lambda$ can acquire mass in general (\textit{i.e.} Zeeman effect). The remaining symmetry is the $M_y \mathcal{T}$ symmetry which relates the cone at $+\Lambda$ to the one at $-\Lambda$:
\begin{equation}
H_{-\Lambda} = (M_y \mathcal{T})H_{+\Lambda}(M_y \mathcal{T})^{-1}= v_x \sigma_y p_x +v_y \sigma_x p_y -\delta v_x p_x +\Delta \sigma_z. 
\end{equation}
To simplify notation, we rotate the basis: $\sigma_x \rightarrow -\sigma_y,\; \sigma_y\rightarrow \sigma_x, \; \sigma_z \rightarrow \sigma_z$ and obtain the following unitarily equivalent Hamiltonian:
\begin{equation}\label{LambdaHam}
H_{\pm \Lambda} = v_x \sigma_x p_x -v_y \sigma_y p_y \pm \delta v_x p_x +\Delta \sigma_z. 
\end{equation}
This is Eq. (1) in the main text, which provides the starting point for our study. In the main text, we have referred to the symmetry that protects the degeneracy of the two valleys as a ``two-fold rotation" or ``inversion" symmetry, but when applied specifically to the surface of TCI, it has a precise meaning as the $M\mathcal{T}$ symmetry. This symmetry is spontaneously broken by Coulomb interaction and results in a valley-polarized state.

\subsection{Massive and tilted Dirac Landau levels}\label{A}
\setcounter{equation}{0}
Assuming valley-polarization, we focus on the one at $+\Lambda$. Under an out-of-plane magnetic field $-B\hat{z}$, in the un-tilted limit, the massive Dirac Hamiltonian in Eq. (\ref{LambdaHam}) can be written as 
\begin{equation}
H_0=\frac{\sqrt{2}\;v}{l_B} \begin{pmatrix}
\lambda & a^\dagger \\
a & -\lambda
\end{pmatrix}
\label{dirac}
\end{equation}
where $ v=\sqrt{v_x v_y}$, magnetic length $l_B=\sqrt{\hbar c/eB}$  and the mass parameter $\lambda = \Delta l_B/(\sqrt{2}v)$. Here, $a^\dagger, a$ are parabolic Landau level raising and lowering operators respectively, and are related to the momentum operators by:
\begin{equation}\label{rescaledmomentum}
\begin{split}
p_x= \frac{1}{l_B}\sqrt{\frac{v_y}{2v_x}} (a^\dagger +&a)\;,\quad p_y =\frac{i}{l_B}\sqrt{\frac{v_x}{2v_y}}(a-a^\dagger)\\ &[a,a^\dagger]=1
\end{split}
\end{equation}

The wavefunctions of the massive Dirac Landau levels and their corresponding energy can then be solved exactly. For the $n$-th Landau level with $n \neq 0$:
\begin{equation}
\begin{split}
&\psi_{n,m} = \frac{1}{\sqrt{1+\gamma_{n}^2}} \begin{pmatrix}
\phi_{\abs{n},m} \\ \gamma_{n} \phi_{\abs{n}-1,m}
\end{pmatrix},\quad E_{n} = s_n \frac{\sqrt{2}\;v}{l_B}\sqrt{\lambda^2 +\abs{n}}
\end{split}
\end{equation}
where
\begin{equation}
\gamma_{n}=\frac{-\lambda +s_n \sqrt{\lambda^2 +\abs{n}\;}}{\sqrt{\abs{n}}}
\end{equation}
Here $s_n=\text{sgn}(n)$, and $\phi_{\abs{n},m}$ are the wavefunctions for a parabolic Landau level in the symmetric gauge with angular momentum $m-\abs{n}$. For the 0-th Dirac Landau level, we have:
\begin{equation}
\psi_{0,m} = \begin{pmatrix}
\phi_{0,m} \\0 
\end{pmatrix}, \quad\; E_0 = \frac{\sqrt{2}\;v}{l_B}\lambda
\end{equation}

When the tilt of Dirac cone $\delta v_x$ is turned on, we can do first order perturbation theory to obtain the approximate eigenstates. To leading order in $\tau = \delta v_x / (2v_x)$, for the $n\neq 0$ massive and tilted Dirac LL, we obtain (up to normalization):
\begin{equation}\label{tiltedDirac}
\psi_{n,m} =
\begin{pmatrix}
\phi_{\abs{n},m} \pm \tau [\alpha_{-1}\phi_{\abs{n}-1,m} +\alpha_1\phi_{\abs{n}+1,m}] \\
\gamma_{n} [\phi_{\abs{n}-1,m} \mp \tau(\alpha_0 \phi_{\abs{n},m} +\alpha_{-2}\phi_{\abs{n}-2,m} )]
\end{pmatrix}
\end{equation}
\noindent where
\begin{equation}\label{alpha}
\small{\begin{split}
&\alpha_{-1} = \frac{(2\abs{n}-1)\sqrt{\lambda^2+\abs{n}} \pm \lambda}{\sqrt{\abs{n}}}\;,\;\alpha_1= - 2\sqrt{\abs{n}+1}\sqrt{\lambda^2+\abs{n}}\\
&\alpha_0 = \frac{(2\abs{n}+1)\sqrt{\lambda^2+\abs{n}} \pm \lambda}{\sqrt{\abs{n}}}\;,\;\alpha_{-2}= - 2\sqrt{\abs{n}-1}\sqrt{\lambda^2+\abs{n}}
\end{split}}
\end{equation}
As for the massive and tilted $0$-th Dirac LL:
\begin{equation}
\psi_{0,m} = \begin{pmatrix}
\phi_{0,m}-2\tau \lambda \phi_{1,m} \\ -\tau \phi_{0,m}
\end{pmatrix}
\end{equation}
These expressions allow us to calculate dipole moments, and energy shifts under the influence of impurity, straightforwardly. 

\subsection{Number of impurity states for massive and tilted Dirac cones}\label{B}
\setcounter{equation}{0}
Here we demonstrate that there are only two states that have probability amplitudes at the impurity site, and which therefore are split from the Landau level, even in the presence of perturbations in mass and tilt of the Dirac cone. 

We consider a delta-function impurity $H_{imp}=V_0 l_B^2 \delta(\vec{x})$. Upon projection to a specific Landau level, the impurity Hamiltonian has matrix elements:
\begin{equation}
\bra{n,m}H_{imp}\ket{n,m'} = V_0 l_B^2 \Psi^\dagger_{n,m}\Psi_{n,m'}
\end{equation}
where we have defined $\Psi_{n,m} \equiv \psi_{n,m} (\vec{0})$, $i.e.$ the amplitude of the Landau level orbital at the impurity site. To the first order in tilt $\tau$, the Dirac Landau level is found in Eq. (\ref{tiltedDirac}). To prove our claim in full generality, let us assume we have carried out a $k$-th order perturbation theory in $\tau$, so that the $n$-th tilted Dirac Landau level $\psi_{n,m}$ is expressed in terms of $\phi_{p,m}$ with $p=\abs{n}-k-1,...,\abs{n}+k$. The only states that are relevant to our impurity problem are those that have non-vanishing probability amplitudes at the impurity site, which correspond to those $\psi_{n,m}$ with $m=\abs{n}-k-1,...,\abs{n}+k$. We thus study the degenerate perturbation theory within this subspace, and consider linear combinations of $\Psi_{n,m}$:
\begin{equation}\label{eq19}
\Phi=r_1\Psi_{n,\abs{n}-k-1}+r_2\Psi_{n,\abs{n}-k}+...+r_{2k+2}\Psi_{n,\abs{n}+k}
\end{equation}
If there is a choice of $(r_1,r_2,...,r_{2k+2})$ such that $\Phi = (0,0)^T$, the corresponding linear combination of intra-Landau level orbitals are guaranteed to diagonalize the impurity Hamiltonian and thus remain at the same energy as the Landau level in the absence of impurity. Below, we argue that there are $2k$ such solutions.

Denote $\Psi_{n,m} = (\psi_{m}^\uparrow,\; \psi_{m}^\downarrow)^T$. Only the intra-Landau level index $m$ is made explicit here. Notice that $\psi_{m}^\uparrow$ and $\psi_{m}^\downarrow$ are both real or both imaginary. This is because each of them is proportional to the wavefunction of parabolic Landau level $\phi_{m,m}$ evaluated at the origin, which is real when $m$ is even and is imaginary when $m$ is odd. Redefining $i\Psi_{n,m} \mapsto \Psi_{n,m}$ for odd $m$, Eq. (\ref{eq19}) with $\Phi=(0,0)^T$ becomes a set of simultaneous equations for \textit{real} unknowns $r_i$. Setting $r_{2k+2} =1$ without loss of generality, we reach the following set of equations for $r_i \in \mathbb{R}$:
\begin{equation}\label{eq20}
\begin{cases}
\small{r_1 \psi^{\uparrow}_{n-k-1} + r_2 \psi^{\uparrow}_{n-k}+ ... + r_{2k+1} \psi^{\uparrow}_{n+k-1} = -\psi^{\uparrow}_{n+k}} \\
\small{r_1 \psi^{\downarrow}_{n-k-1} + r_2 \psi^{\downarrow}_{n-k}+ ... +r_{2k+1} \psi^{\downarrow}_{n+k-1} = -\psi^{\downarrow}_{n+k}}
\end{cases}
\end{equation}

With $2k+1$ unknowns and only two linear equations, there are in general $2k$ linearly independent solutions, leading to $2k$ states that have vanishing amplitudes at the impurity site. Since we start with a $(2k+2)$-dimensional subspace, only $2k+2-2k=2$ states are allowed to have non-vanishing amplitudes at the origin. These are the two impurity bound states whose energy are split from the bulk Landau level, and are the ones employed in our construction of impurity dipole moment in the main text.

The above argument also works for the 0-th Landau level. However, only one impurity state is significantly shifted away from the bulk Landau level, while the shift of the second impurity state is minuscule (controlled by the size of the tilt), so practically, in the quantum Hall ferroelectric system that we consider, only one impurity state can be probed in this special case.

\subsection{Choice of Parameters}\label{C}
In the main text, we use the following parameters to study the quantum Hall ferroelectrics in topological crystalline insulator Sn$_{1-x}$Pb$_x$(Te,Se):
\begin{equation}\label{C1}
\tau=0.1,\quad \lambda=0 ,\quad v_x/v_y=1.6
\end{equation}
Here we explain why these values match with the low-energy physics of the system obtained either from experiments or \textit{ab initio} calculations.

The tilting effect of Dirac cones (at $\bar{\Lambda}$) has been observed in the ARPES measurements by Tanaka \textit{et al.} \cite{supp15}, from which we estimate the tilting parameter to be $\tau=\delta v_x/(2v_x)=0.1$. The acquisition of mass in topological crystalline insulators was observed by Okada \textit{et al.} \cite{supp5}. Under a ferroelectric distortion, two of the four surface Dirac cones were measured to obtain mass of about $\Delta=10 $ meV, and this would correspond to $\lambda=\Delta l_B/(\sqrt{2}v)\approx 0.5$. But notice that these two massive cones are located near $\bar{X}_1$ (see Fig. \ref{BZ}), which are not symmetry-related in our quantum Hall setting. In our study we are instead focusing on the Dirac cones near $\bar{X}_2$, whose degeneracy is symmetry-protected as explained in Sec. \ref{AA}. These cones can acquire a mass via the Zeeman effect, but the experiment by Okada \textit{et al.} \cite{supp5} suggests that it is too small to be observed. A rough estimate with the vacuum Zeeman effect would give $\lambda \sim 0.01$, and thus in the main text we decide to assume $\lambda=0$. The values of $v_x$ and $v_y$ have been obtained by Liu \textit{et al.} by fitting with \textit{ab initio} calculations \cite{supp12}. For the effective Dirac Hamiltonian (around $\bar{\Lambda}$) that we are considering, $v_x=1.3$ eV$\AA$ and $v_y=0.83$ eV$\AA$. Thus, the anisotropy $v_x/v_y=1.6$. 

\subsection{Relation between the local impurity dipole and the bulk adiabatic dipole}\label{CD}
Here we present more details of the derivations of Eqs.(4) and (6) in the main text. \\

Let us begin with the bulk adiabatic dipole moment introduced in the modern theory of polarization \cite{supp1}. Since the tilt of Dirac cone in our model breaks inversion symmetry in the $p_x$-direction, electric polarization is non-vanishing only in the $y$-direction and we thus focus only on that component. Following the conventional Berry phase approach, we have \cite{supp2}:
\begin{equation}\label{CD1}
D_y = ie\frac{l^2_B}{L_x}\int_{0}^{L_x/l^2_B}dk_y\;\bra{u_{k_y}}\partial_{k_y}\ket{u_{k_y}}
\end{equation}
where $\ket{u_{k_y}}$ is the Bloch wavefunction in a gauge which is invariant under translation in $y$-direction. In this gauge, the complete wavefunction is $\psi_{k_y}(x,y) = \frac{e^{ik_y y}}{\sqrt{L_y}}u_{k_y}(x)$. Making use of $u_{k_y}(x) = u_0(x-k_y \_B^2)$, one can recast Eq.(\ref{CD1}) into the following form:
\begin{equation}
\begin{split}
D_y&=-i e l^2_B \int d^2 r\;\psi^*_{k_y}(x,y) \partial_x \psi_{k_y}(x,y) \\
&= e l^2_B \bra{\psi_{k_y}}p_x\ket{\psi_{k_y}}
\end{split}
\end{equation}
Expressing the momentum operator in terms of inter-Landau level ladder operators as in Eq. (\ref{rescaledmomentum}), and acting it on the first-order perturbed Landau orbitals in Eq. (\ref{tiltedDirac}), one would obtain Eq. (4) of the main text:
\begin{equation}\label{CD2}
\begin{split}
&{\bf D}_{n} =  {\tilde s}_{n} \sqrt{2}\;\tau\; e \;l_B\;\left(\frac{2\lambda^2+3|n|}{\sqrt{\lambda^2+|n|}}\right)\sqrt{\frac{v_y}{v_x}} \hat{{\bf y}},
\end{split}
\end{equation}
where $\tilde{s}_n = \text{sgn}(n)$ (and $\tilde{s}_0=1$). 
\\

\begin{figure}[H]
\centering
\includegraphics[width=1\textwidth, height=4.3cm]{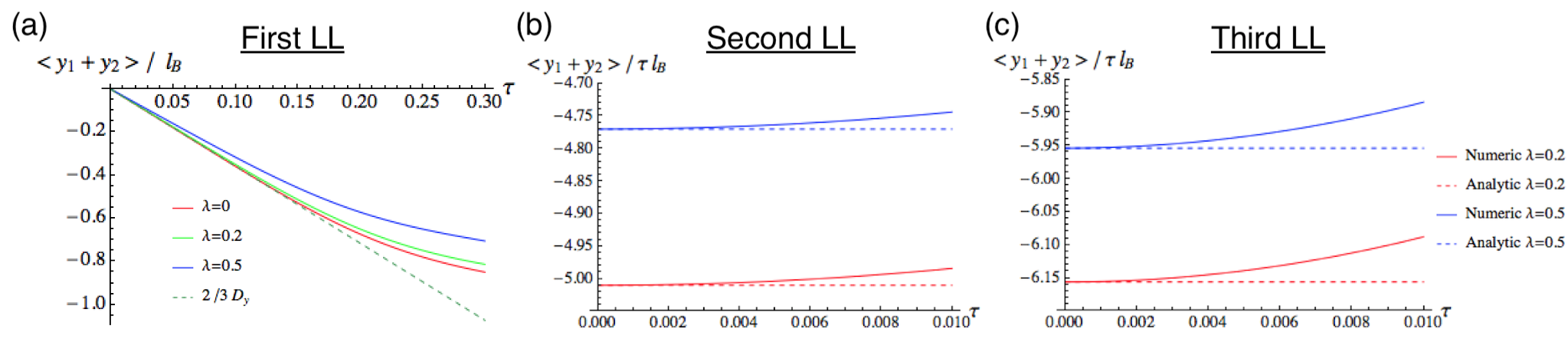}
\caption{Numerical checks of the relation between impurity dipole and bulk adiabatic dipole. (a) Impurity dipole $D^{\rm imp}_y$ (with $e$ set to 1) as a function of tilt $\tau$, for the first Dirac Landau level in the valence band with various mass $\lambda$. The dashed line is given by the analytic result in Eq. (\ref{CD6}). (b) and (c): $D^{\rm}_y/\tau$ as a function of tilt $\tau$ for the second and third Landau levels in the valence band respectively. Solid lines represent results from solving the impurity states numerically, while dashed lines are generated using the analytic formula in Eq. (\ref{CD7}). For small enough tilt such that the first-order perturbation theory is sufficient, it is shown that the analytic result complies with the numerical calculation. Note that the figures here show the analysis for Landau levels in the valence band ($n<0$), which differ from the situation in the conduction band ($n>0$) simply by a minus sign in the dipole.}
\label{impuritydipole}
\end{figure}

Next we turn to the local impurity dipole moment, which is defined as 
\begin{equation}\label{CD3}
{D}^{\rm imp}_y= e(\bra{\tilde{\psi}_1}\hat{y}\ket{\tilde{\psi}_1}+ \bra{\tilde{\psi}_2}\hat{y}\ket{\tilde{\psi}_2}),
\end{equation}
where $\tilde{\psi}_{1,2}$ are the two impurity states bound to the delta-potential defect. Expressing the position operator $\hat{y}$ in terms of the guiding-center operator and the momentum operator: $\hat{y}=\hat{R}_y-l^2_B \hat{p}_x$, it follows that:
\begin{equation}\label{CD4}
D^{\rm imp}_y = e(\bra{\tilde{\psi}_1}\hat{R}_y\ket{\tilde{\psi}_1}+\bra{\tilde{\psi}_2}\hat{R}_y\ket{\tilde{\psi}_2}) - 2D_y.
\end{equation}
Notice that the guiding-center operator can be expressed in terms of \textit{intra}-Landau level ladder operators: $\hat{R}_y = il_B(b^\dagger-b)/\sqrt{2} $. To proceed analytically, we first consider the massless limit and to leading order in the tilt of Dirac cone. The impurity states for the $n$-th Landau level have the following expressions:
\begin{subequations}\label{CD5}
\begin{align}
\tilde{\psi}_1&=-s_n\frac{i\tau\alpha_{-2}}{\sqrt{2}}\psi_{n,n-2}+(-\frac{1}{\sqrt{2}}+\frac{\tau n}{\sqrt{2}})\psi_{n,n-1}+s_ni(\frac{1}{\sqrt{2}}+\frac{\tau n}{\sqrt{2}})\psi_{n,n}-\frac{\tau\alpha_1}{\sqrt{2}} \psi_{n,n+1}\\
\tilde{\psi}_2&=s_n\frac{i\tau\alpha_{-2}}{\sqrt{2}}\psi_{n,n-2}+(\frac{1}{\sqrt{2}}+\frac{\tau n}{\sqrt{2}})\psi_{n,n-1}+s_ni(\frac{1}{\sqrt{2}}-\frac{\tau n}{\sqrt{2}})\psi_{n,n}-\frac{\tau\alpha_1}{\sqrt{2}} \psi_{n,n+1}
\end{align}
\end{subequations}
where $s_n = \text{sgn}(n)$. Note that $\psi_{n,m}$ and $\alpha_i$ are defined in Eqs. (\ref{tiltedDirac}) and (\ref{alpha}) respectively. Combining Eqs. (\ref{CD4}) and (\ref{CD5}), we obtain:
\begin{equation}\label{CD6}
{\bf D}_{n}^{\rm imp} = \frac{2}{3} {\bf D}_{n}
\end{equation}
in the massless limit. For the massive case, the algebra becomes complicated without specifying an explicit LL index. For $n=1$ and $n=2$, we have obtained explicit analytic expressions for the impurity states and evaluated the impurity dipole. The result suggests that 
\begin{equation}\label{CD7}
{\bf D}_{n}^{\rm imp} = \frac{2|n|}{3|n|+2\lambda^2} {\bf D}_{n}. 
\end{equation}
The validity of this expression is further checked numerically for higher Landau levels (see Fig. \ref{impuritydipole}). It is also worth pointing out that for higher Landau levels, we are usually in the regime where $\lambda\ll \sqrt{\abs{n}}$, so the relation in Eq. (\ref{CD6}) for the massless limit is a good approximation. 

\subsection{Differences in the two notions of electric dipole moment}\label{D}
\setcounter{equation}{0}
To further clarify the difference between the adiabatic bulk dipole moment (following the modern theory of polarization) and the impurity dipole moment $\textbf{D}^{\text{imp}}$ introduced in this paper, we consider a toy model with a parabolic dispersion:
\begin{equation}\label{D1}
H=\frac{(p_x-a_x)^2}{2m_x}+\frac{p_y^2}{2m_y}
\end{equation}
The parameter $a_x$ plays a similar role as the tilt $\delta v_x$ in the Dirac Hamiltonian.

Now, apply a magnetic field $B\hat{z}$ on the system. Denote a Landau orbital as $\ket{\psi}$ for $a_x$=0, and the corresponding Landau orbital when $a_x$ is tuned from zero to some finite value as $\tilde{\ket{\psi}}$. According to the polarization theory based on Berry phase, the difference of polarization between these two Landau orbitals is:
\begin{equation}\label{D2}
\begin{split}
\Delta D_y &= -\abs{e}l^2[\tilde{\bra{\psi}}\;p_x\;\tilde{\ket{\psi}}-\bra{\psi}p_x\ket{\psi}]\\
&=-\abs{e}l^2[ \tilde{\bra{\psi}}\;\tilde{p}_x+a_x\;\tilde{\ket{\psi}}-\bra{\psi}p_x\ket{\psi}]\\
&=-\abs{e}l^2[\tilde{\bra{\psi}}\;\tilde{p}_x\;\tilde{\ket{\psi}}-\bra{\psi}p_x\ket{\psi}+a_x]\\
&=-\frac{a_x}{\abs{B}}
\end{split}
\end{equation}
The last equality is obtained because $\tilde{p}_x=p_x-a_x$ is just a gauge transformation, while the expectation value $\bra{\psi}p_x\ket{\psi}$ should be gauge-invariant.

However, this dipole moment does not reflect the inversion asymmetry of the Landau orbital. In this example, there is simply no inversion asymmetry to begin with, and this can be verified if one examine ${\bf D}^{\rm imp} = e\langle \phi| {\bf  r}|\phi\rangle$, for the Landau orbital bound to a delta-potential impurity. By a proper gauge transformation, one can move the center of unperturbed Landau orbitals to the impurity site, irrespective of what $a_x$ is. After all, the presence of $a_x$ can be viewed as a gauge-transformation. In the presence of a delta-function impurity, only one state in each Landau level is bound to the impurity. That is the state $\phi_{n,n}$, which has a non-zero amplitude at the origin where the impurity sits.  As this state is inversion symmetric, and the perturbation (i.e. the delta potential) preserves this symmetry, the bound state should also be inversion symmetric. Thus $\bf D^{\rm imp}=0$. 

In this extreme example, which can be considered as the parabolic limit ($\lambda \rightarrow \infty$) of the Dirac Hamiltonian, $\bf D$ measures solely the effect of Landau orbital displacement, which cannot be detected in a quantum Hall system due to edge screening. On the other hand, $\bf D^{\rm imp}$ measures only the inversion asymmetry of Landau orbitals, and therefore gives a local experimental signature for quantum Hall ferroelectrics. \\

\subsection{Numerical Setup of Exact Diagonalization}\label{E}
\setcounter{equation}{0}
\subsubsection{Anisotropic parabolic dispersions}

To exact-diagonalize the Hamiltonian with Coulomb interaction, one has to project the Coulomb term onto the Landau orbitals. In the main text, we deal with Landau levels arising from the tilted Dirac cones dispersion, while in this appendix we will also consider the case with parabolic dispersion. The parabolic case is the cornerstone for case with tilted Dirac cones dispersion, since the Dirac Landau orbitals are spinors consisting of parabolic Landau orbitals. The parabolic dispersion Hamiltonian is:
\begin{equation}\label{parabolic}
H=\frac{1}{2m^*}p_a g_{ab}p_b=\frac{1}{l_B^2m^*}(a^\dagger a+\frac{1}{2})
\end{equation}
where $p = \nabla/i-eA$, $g = Q^T S^2 Q$ is a $2\times 2$ tensor,  $Q\in SO(2)$ describes the rotation around principal axes in real space, the valleys we are interested in are vertical oriented, so we can set the real space axes along the principal axes of rotation, thus $Q=I$, and simply $g=S^2$. $S = diag\{(m_x/m_y )^{1/4}(m_y /m_x )^{1/4} \} $ is the mass tensor for the valley , effective mass $m^* = (m_x m_y )^{1/2}$. We introduce the mass ratio: $\alpha=m_x/m_y$ that specifies aspect ratio of the valley. The rescaled momenta along the principal axes of the tensor $\pi_a = S_{ab}p_b$ satisfy:
\begin{equation}
[\pi_a,\pi_b]=i l_B^{-2}\epsilon_{ab}
\end{equation}
and the LL lowering operator is:
\begin{equation}\label{momentum}
a=\frac{l_B}{\sqrt{2}}(\pi_x+i\pi_y),\quad [a,a^\dagger]=1
\end{equation}
Numerically, the electrons are on the 2D surface of torus, $L_x(L_y)$ represents the circumference of the torus along $x(y)$ direction and they satisfy relation $L_xL_y=2\pi N_0$, $N_0$ represents the number of orbitals for each valley.

Choosing the Landau gauge, $\vec{A}=(0,x)B$, the wavefunction of LL orbital is expressed as:\\
\begin{equation}
\begin{split}
\phi_{n,j}^\alpha&(r)=\left(\frac{2\pi}{L_yl_B}\right)^{1/2}\Sigma_{k=-\infty}^{+\infty}\overline{H_n}\left[\frac{x-k L_x-X_j}{\alpha^{1/4}l_B}\right]\times\exp[i(X_j+k L_x)y/l_B^2-(X_j+k L_x-x)^2/(2 \alpha^{1/2}l_B^2)]\\
\end{split}
\end{equation}
where $X_j=\frac{2\pi l_0^2 j}{L_y}$ to fulfill the periodic boundary condition, $\overline{H_n}$ is the physicist's Hermite polynomial that has been normalized so that:
\begin{equation}
\int^{+\infty}_{-\infty} \left(\overline{H_n}(x)\right)^2 e^{-x^2}dx=1
\end{equation}

With $\phi^\alpha_{n,j}(r)$ normalized as $\int_0^{L_y}dy\int_0^{L_x}dx|\phi_{n,j}^\alpha(r)|^2=2\pi$, when a Landau level is completely filled and thus the electron density is uniformly distributed, $\int_0^{L_y}dy\int_0^{L_x}dx\sum_j |\phi_{n,j}^\alpha(r)|^2=2\pi N_0=L_x L_y$ would then imply $\sum_j |\phi_{n,j}^\alpha(r)|^2=1$.

Next, we define $f_{nm}$ as the form factor for the parabolic Landau levels calculated in the Landau gauge:
\begin{equation}\label{factor}
\begin{split}
f_{nm}(\textbf{q}^\alpha)  &=\langle n,\alpha|e^{il_B^2\textbf{q}^\alpha\cdot\bm{\pi}}|m,\alpha\rangle \\
&=e^{-l_B^2(q_y^\alpha)^2/4}\int_{-\infty}^{+\infty}\overline{H_m}(x-\frac{q^\alpha_y}{2})\overline{H_n}(x+\frac{q^\alpha_y}{2})e^{-x^2}e^{iq^\alpha_x x}dx
\end{split}
\end{equation}
where the wavevector $\textbf{q}^\alpha$ is not the natural wavevector $\textbf{q}$ but rotated as
\begin{equation}
\textbf{q}^\alpha = -S^{-1}\epsilon\textbf{q}
\end{equation}
where $\epsilon$ is the rank-2 levi-civita symbol, $Q$ and $S$ are the matrices associated with the mass ratio $\alpha$. This definition will become clear later when we project the electron interaction on the LLs.
\subsubsection{Tilted Dirac cone dispersion}
The massless Dirac Hamiltonian is just Eq. (\ref{dirac}) with $\lambda=0$. 
Similar to the case with a parabolic Hamiltonian in Eq. (\ref{parabolic}), here we would define $S=diag\{ (v_x/v_y)^{1/2},(v_y/v_x)^{1/2}\}$ and rescale the momentum by $\pi_a = S_{ab}p_b$, which explains Eq. (\ref{rescaledmomentum}). One can relate the mass ratio $\alpha$ in the anisotropic parabolic dispersion and velocity ratio $r=v_x/v_y$ in the Dirac dispersion as: $\alpha=r^2$.

Tilting of the Dirac cone along the $x$-direction is described by the following perturbation:
\begin{equation}
H_1=\delta v_x p_x=\delta v_x\sqrt{\frac{v_y}{v_x}}\frac{(a+a^\dagger)}{\sqrt{2}l_B}=\tau\frac{\sqrt{2}v}{l_B}(a+a^\dagger)
\end{equation}
where $\tau=\delta v_x/(2 v_x)$. Using the general expression for the tilted Dirac LL in Eq. (\ref{tiltedDirac}), we have the following expression for the $n=+3$ Dirac Landau level:
\begin{equation}
\begin{split}
|+3,\tau\rangle=\frac{1}{\sqrt{2}}\left(\begin{matrix}|3\rangle+\tau(-4\sqrt{3}|4\rangle+5|2\rangle)\\ |2\rangle+\tau(2\sqrt{6}|1\rangle-7|3\rangle)\end{matrix}\right)
\end{split}
\end{equation}
Here, for simplicity, we have suppressed the intra-Landau level indices and the mass ratio $\alpha$ that would label the parabolic Landau orbitals. The form factor for the Dirac Landau level is then obtained as follows:
\begin{equation}\label{Dfactor}
\begin{split}
F^3(\textbf{q}^\alpha,\tau)&=\langle +3,\tau|e^{il_B^2\textbf{q}^\alpha\cdot\bm{\pi}}|+3,\tau\rangle \\ &=\frac{1}{2}[f_{33}+f_{22}-2\tau(f_{32}+f_{23})-4\sqrt{3}\tau(f_{34}+f_{43})+2\sqrt{6}\tau(f_{12}+f_{21})]
\end{split}
\end{equation}
where $f_{nm}$ is the form factor for the parabolic Landau levels (Eq. \ref{factor}).

\subsubsection{Impurity potential}
The impurity potential is $U(\textbf{r})=V_0l_B^2\delta(\textbf{r})$.
The matrix elements of impurity potential projected to the $n$-th and $m$-th parabolic Landau levels are:
\begin{equation}
\begin{split}
U&^\alpha_{j_1,j_2,n,m}
=V_0\frac{2\pi l_B}{L_y}\Sigma_{l=-\infty}^{+\infty}\Sigma_{k=-\infty}^{+\infty}\overline{H_n}\left[\frac{X_{j_1}+lL_x}{l_B\alpha^{1/4}}\right]\times\overline{H_m}\left[\frac{X_{j_2}+kL_x}{l_B\alpha^{1/4}}\right]e^{-\frac{(X_{j_1}+l L_x)^2+(X_{j_2}+k L_x)^2}{2l_B^2\sqrt{\alpha}}}
\end{split}
\end{equation}
In the parabolic dispersion case, one only need to consider the case $n=m$, and in the main text we focus on the lowest Landau level, so $n=m=0$; on the other hand, in the tilted Dirac case there exist non-trivial terms with $n\neq m$, the impurity matrix elements $\langle +3,\tau|\hat{U}(\textbf{r})|+3,\tau\rangle_{j_1,j_2}$ are linear combinations of $U_{j_1,j_2,n,m}$ with $n,m=1,2,3,4$, which is similar to the form factor in Eq. (\ref{Dfactor}).

\subsubsection{Coulomb interaction}
The Coulomb interaction in a finite system has the form
\begin{equation}
V(\textbf{r})=\frac{1}{L_x L_y}\sum_\textbf{q} V(q) e^{i \textbf{q}\cdot \textbf{r}}
\end{equation}
where $V(q)=\frac{2\pi e^2}{\epsilon q}$, for finite size torus with the circumference $L_x$ and $L_y$. Here $\textbf{q}=(\frac{2\pi s}{L_x},\frac{2\pi t}{L_y})$ takes discrete values to ensure the periodicity.

The projected Coulomb interaction between two electrons in the valleys $i$ and $j$ ($i,j$ can either be the same valley or two different valleys) into the $n$-th Landau level has the form:
\begin{equation}
\begin{split}
P_n &V(\textbf{r}_i-\textbf{r}_j)P_n=\frac{1}{L_xL_y}\sum_\textbf{q} V(\textbf{q}) F^n_i(\textbf{q}_i) F_j^{n}(\textbf{q}_j)^*e^{i \textbf{q}\cdot (\textbf{R}_i-\textbf{R}_j)}
\end{split}
\end{equation}
Here we have introduced the guiding center operator $\textbf{R}_{i}$ for valley $i$, which is related to the position operator as follows:
\begin{equation}
\textbf{r}_i \equiv \textbf{R}_i-l_B^2\epsilon\textbf{p}_i =  \textbf{R}_i-l_B^2\epsilon S_i^{-1}\bm{\pi}_i 
\end{equation}
where $\epsilon$ is the rank-2 levi-civita symbol and $S_i$ is the $S$ tensor associated to valley $i$, which has been defined earlier for both parabolic and Dirac dispersions. Accordingly, wavevector $\textbf{q}_i$ is defined as:
\begin{equation}
\textbf{q}_i = -S_i^{-1}\epsilon\textbf{q}
\end{equation}
For the numerical results presented in the main text, the valleys have the same velocity ratio $r$ (or mass ratio $\alpha=r^2$) and opposite $\tau$. Thus we have $\textbf{q}_i=\textbf{q}_j=\textbf{q}^\alpha, F_{i/j}^n(\textbf{q}^\alpha)=F^3(\textbf{q}^\alpha,\pm\tau)$.  While for the numerics to be presented in the next section for anisotropic parabolic dispersion at $n=0$ LL, different valleys have different mass ratio $\alpha$ and $\beta$, where $\beta=\frac{1}{\alpha}$ for the two orthogonal-orientated valleys of interest. There we have $\textbf{q}_{i/j}=\textbf{q}^{\alpha/\beta}$, and $F^n_{i/j}=f_{00}$.\\

\subsection{Quantum Hall Nematics with Anisotropic Parabolic Dispersions}\label{F}
\begin{figure}[H]
\centering
\includegraphics[width=0.4\textwidth]{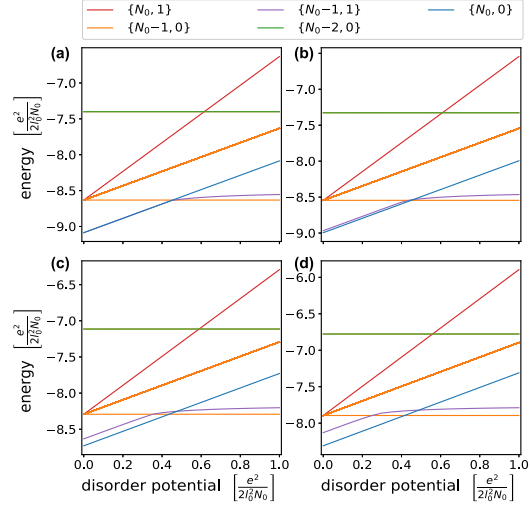}
\caption{The energy spectra with increasing impurity potentials, as indicated in the legend, blue lines represent $\{N_0,0\}$, red lines $\{N_0,1\}$, orange lines $\{N_0-1,0\}$, purple lines $\{N_0-1,1\}$ and green lines $\{N_0-2,0\}$.  The mass ratios($\alpha=m_x/m_y$) in panels (a),(b),(c),(d) are $1,2,4,8$, respectively. The orbital number $N_0=20$. }
\label{paraspectrum}
\end{figure}

\begin{figure}[H]
\centering
\includegraphics[width=0.4\textwidth]{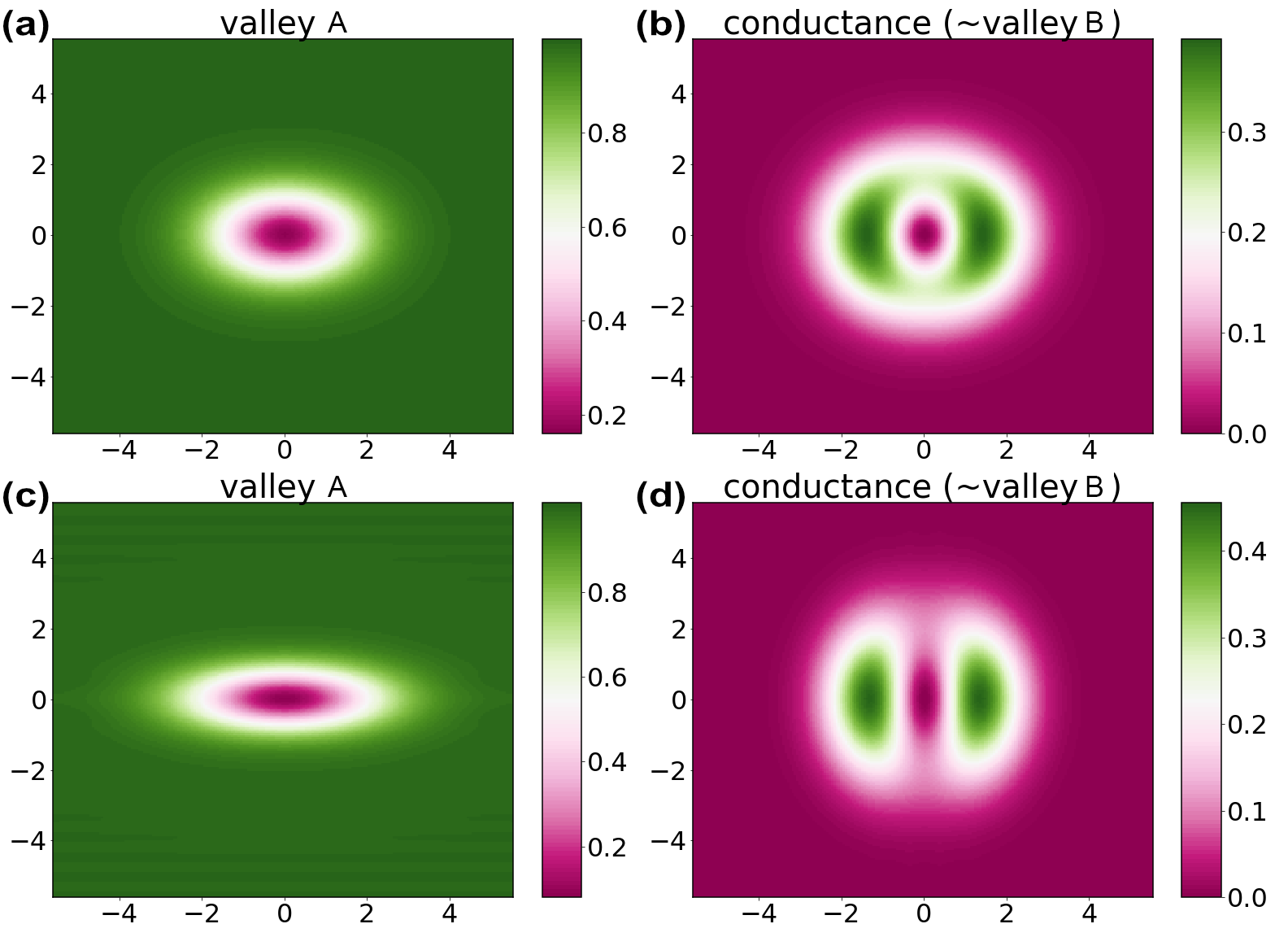}
\caption{The tunneling matrix elements from the ground state $\{N_0-1,0\}$ to the excitonic state $\{N_0-1,1\}$ for different mass ratio: $\alpha=2$ in (a, b) and $\alpha=8$ in (c,d),  which are proportional to the differential conductance obtained by direct STM measurements. The strength of impurity potential is set to be $0.6\frac{e^2}{2l_B^2N_0}$ and the length scale is in the unit of $l_B$.}
\label{paratunneling}
\end{figure}

 After considering electron-electron interaction in ferroelectric states in the main text, here we illustrate a simpler scenario where the anisotropic parabolic dispersion is used so that the impurity only hosts a single bound state. The two valleys A and B are parabolic dispersive with the same aspect ratio, but vertical elliptical axes, meaning that if we choose the principal axes along the same direction for two valleys, there mass ratio will satisfy $\alpha=\frac{1}{\beta}$.  A smaller system size with $N_0=20$ single-valley orbitals is enough to demonstrate this case. The corresponding energy spectra with disorder are shown in Fig. \ref{paraspectrum}, and some representative tunneling density profiles are shown in Fig. \ref{paratunneling}, with various mass ratios. Again, just like what happens in the ferroelectric state around an impurity, when the impurity potential is larger than a certain threshold a quasihole state becomes the new ground state. Adding an electron to this state would lead to an exciton state, and the resulting density profile can be captured by STM measurements.

\begin{thebibliography}{99}
\bibitem{23} For simplicity we neglect inter-valley scattering although it has been recently shown that certain detailed aspects of the states near the impurity require understanding the intervalley scattering properties~\cite{20}.
\end{thebibliography}
\end{document}